\documentclass[aps,prl,showpacs,reprint,superscriptaddress]{revtex4-1}

\usepackage{graphicx}
\usepackage{braket}
\usepackage{amsmath}
\usepackage{amssymb}
\usepackage{color}

\newcommand{\vp}{\varphi}

\begin{document}


\title{Effective Confining Potential of Quantum States in Disordered Media}

\author{Douglas N. Arnold}
\email[]{arnold@umn.edu}
\affiliation{School of Mathematics, University of Minnesota, Minneapolis, Minnesota 55455, USA}

\author{Guy David}
\email[]{guy.david@math.u-psud.fr}
\affiliation{Universit\'e Paris-Sud, Laboratoire de Math\'ematiques, CNRS, UMR 8658, Orsay F-91405, France}

\author{David Jerison}
\email[]{jerison@math.mit.edu}
\affiliation{Mathematics Department, Massachusetts Institute of Technology, Cambridge, Massachusetts 02139, USA}

\author{Svitlana Mayboroda}
\email[]{svitlana@math.umn.edu}
\affiliation{School of Mathematics, University of Minnesota, Minneapolis, Minnesota 55455, USA}

\author{Marcel Filoche}
\email[]{marcel.filoche@polytechnique.edu}
\affiliation{Physique de la Mati\`ere Condens\'ee, Ecole Polytechnique, CNRS, Palaiseau F-91128, France}

\date{\today}

\begin{abstract}
The amplitude of localized quantum states in random or disordered media may exhibit long-range exponential decay. We present here a theory that unveils the existence of an effective potential which finely governs the confinement of these states. In this picture, the boundaries of the localization subregions for low energy eigenfunctions correspond to the barriers of this effective potential, and the long-range exponential decay characteristic of Anderson localization is explained as the consequence of multiple tunneling in the dense network of barriers created by this effective potential. Finally, we show that the Weyl's formula based on this potential turns out to be a remarkable approximation of the density of states for a large variety of one-dimensional systems, periodic or random.
\end{abstract}


\maketitle


Despite more than 50~years of research, many questions on the exact mechanism of Anderson localization still remain open~\cite{Anderson1958,Abrahams2001,Evers2008,Lagendijk2009}. One of the most puzzling aspects of this phenomenon is the strong spatial confinement of the one-particle quantum states, i.e., the exponential decay of the wave amplitude at long range in the absence of any confining potential~{\color{black}\cite{John1987,Billy2008,Roati2008,Riboli2011}}. In the interpretation due to Anderson, this decay comes from the destructive interferences between waves traveling from an initial source along different propagation pathways in the disordered potential~{\color{black}\cite{Akkermans1988,Kuhn2005}}. At distances from the origin much larger than the correlation length, the waves statistically almost cancel each other out, leading to an exponential decay of the amplitude. However, this statistical result says nothing about the detailed nature of the decay. Does it occur smoothly, in a continuous way, when the distance goes to infinity? Or are there specific places where a transition between constructive and destructive interference can be observed?

A recent theory has shown that the precise spatial location of such quantum states in a potential $V(\vec{r})$ can be predicted using the solution $u(\vec{r})$ of a simple associated Dirichlet problem, called the \emph{localization landscape}~\cite{Filoche2012}. While quantum states and their energies are, respectively, the eigenfunctions and the eigenvalues of the Hamiltonian of the system defined as $\hat{H} = -\displaystyle \frac{\hbar^2}{2m} \Delta + V$, the landscape $u$ is defined as the solution of
\begin{equation}
\label{eq:udef}
\hat{H} u = -\frac{\hbar^2}{2m}~\Delta u + V u = 1~ ,
\end{equation}
the boundary conditions being either Dirichlet, Neumann, or periodic. In this theory, the localization subregions are delimited by the valley lines of the graph of~$u$. This property directly derives from a fundamental inequality satisfied by any eigenfunction~$\psi$ of $\hat{H}$ with eigenvalue~$E$, normalized so that its maximum amplitude is equal to 1:
\begin{equation}
\label{eq:control}
|\psi(\vec{r})| \le E~u(\vec{r})~.
\end{equation}
In other words, the small values of $u$ along its valley lines constrain the amplitude of~$\psi$ to be small along the same lines and, as a consequence, localize low energy eigenfunctions inside the regions enclosed by these lines~\cite{Filoche2012}.

We unveil here a different---and much more powerful role---played by $u$, by showing that the function $W \equiv 1/u$ can in fact be interpreted as a confining potential that is responsible for the exponential decay of the Anderson localized states even far from its main localization subregion. To that end, the original Schr\"odinger equation is transformed by introducing an auxiliary function $\vp$ such that $\psi \equiv u \vp$. Expressing that $\psi$ is an eigenvector of the Hamiltonian leads to
\begin{equation}
\label{eq:vp1}
\left(-\frac{\hbar^2}{2m}~\Delta + V\right) \left(u \vp \right) = E~u \vp~.
\end{equation}
Developing this equation and accounting for the definition of $u$ in Eq.~\eqref{eq:udef} gives
\begin{equation}
\label{eq:vp2}
-\frac{\hbar^2}{2m}~\Delta \vp - 2 \frac{\hbar^2}{2m}~\frac{\nabla u}{u} \cdot \nabla \vp + \frac{1}{u}~\vp = E \vp~.
\end{equation}
The additional first order term proportional to $\nabla \vp$ can be inserted into the second order term which finally yields:
\begin{equation}
\label{eq:vp3}
-\frac{\hbar^2}{2m} \left[\frac{1}{u^2}~\mbox{div}\left(u^2 \nabla \vp \right)\right] + W\vp = E \vp~.
\end{equation}
One can see that the auxiliary function $\vp=\psi/u$ thus obeys a Schr\"odinger-type equation in which the original potential~$V(\vec{r})$ has disappeared. Instead, a new function $W(\vec{r})$ now plays the role of ``effective confining potential.'' One first notices that, since $u$ is a solution of Eq.~\eqref{eq:udef}, $W$ is indeed homogeneous to an energy. Moreover, the valleys of~$u$ which are the boundaries of the localization subregions~\cite{Filoche2012} also correspond to the crest lines of this new potential. These crest lines act as barriers for the auxiliary function~$\vp$.

\begin{figure}
\leftline{
\hspace{0.5mm}
\includegraphics[height=0.18\textwidth]{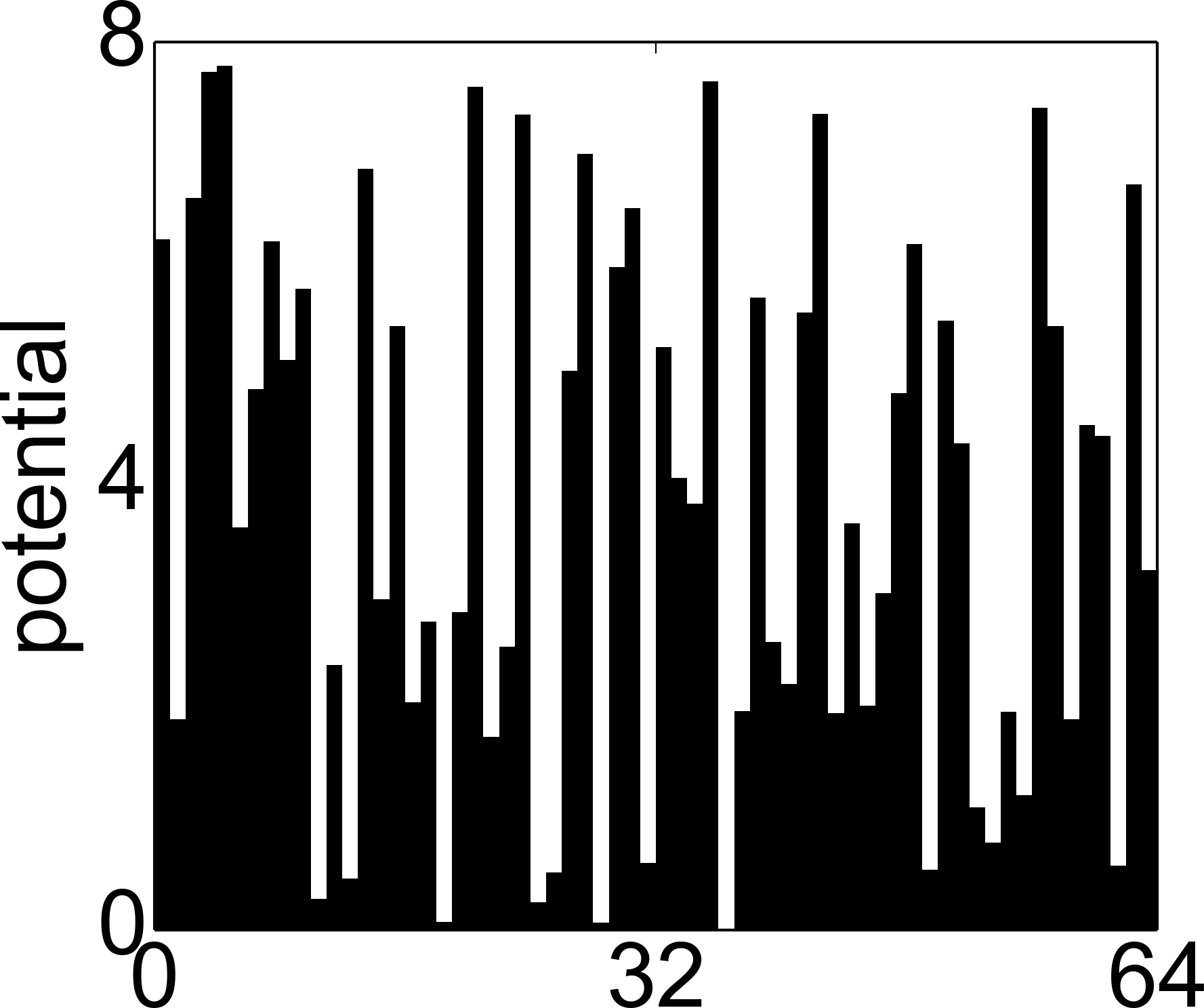}
\hspace{4mm}\includegraphics[height=0.18\textwidth]{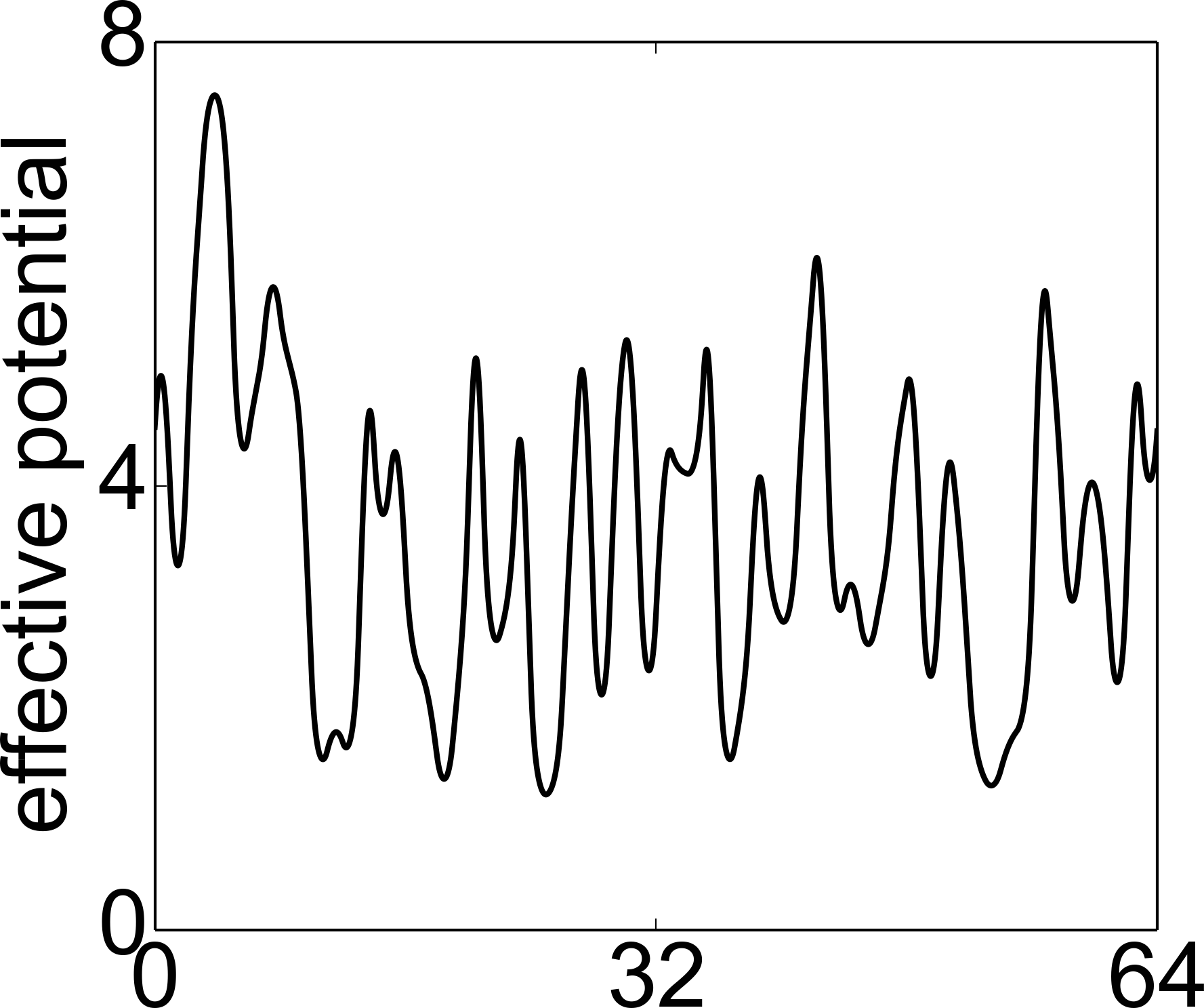}}
\leftline{\hspace{19mm}(a)\hspace{38mm}(b)}
\includegraphics[height=0.16\textwidth]{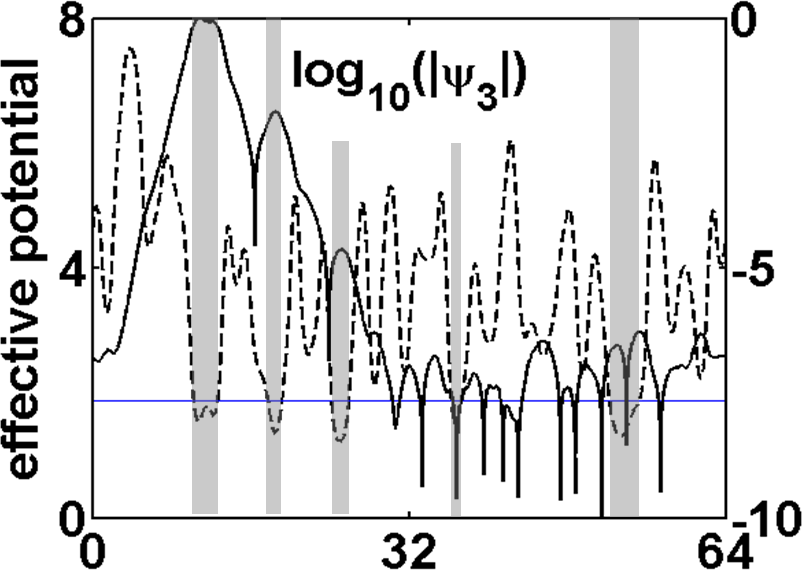}
\hskip 1mm
\includegraphics[height=0.17\textwidth]{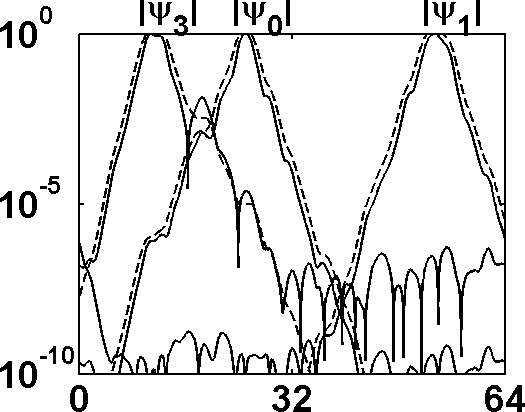}
\leftline{\hspace{19mm}(c)\hspace{38mm}(d)}
\caption{(a) Random piecewise constant potential. (b) Effective confining potential~$W$ computed by solving Eq.~\ref{eq:udef} then taking the reciprocal of~$u$. (c) Probability amplitude of the fourth state $\ket{\psi_3}$ of energy $E_3=1.88$ in a logarithmic scale superimposed on the effective potential~$W$ (the dashed line). The wells of $W$ for this state are defined by the intersections with the horizontal line $W=E_3$ and are outlined in grey. One can see that the decay of the eigenfunction occurs mostly outside these wells, i.e., in the barriers of~$W$, and stops inside these wells. (d) Amplitudes of three different quantum states ($\psi_0$, $\psi_1$, and $\psi_3$) in a logarithmic scale, superimposed with the estimates obtained using the Agmon distance (the dashed lines). These estimates are equal to the right-hand side of Eq.~\eqref{eq:Agmon_ineq}. The Agmon estimates follow in detail the decay of the actual amplitudes down to values smaller than~10$^{-7}$.}\label{fig:u1D}
\end{figure}

The demonstration that~$W$ plays the role of an effective potential derives from the following equality satisfied by any quantum state~$\ket{\psi}$:
\begin{equation}
\label{eq:Agmon}
\braket{\psi|\hat{H}|\psi} = \braket{u~\hat{p} \left(\frac{\psi}{u}\right) | u~\hat{p} \left(\frac{\psi}{u}\right) } ~+~ \braket{\psi|\hat{W}|\psi}~.
\end{equation}
The detailed proof of this equality is given in the Supplemental Material~\cite{SM_PRL2016}. It not only shows that, regardless of its kinetic energy, the energy $E$ of a quantum state~$\ket{\psi}$ can never be smaller than the one it would have in a potential~$W\left(\vec{r}\right)$, but also shows that the difference $(W-E)$ can be used to build an Agmon distance $\rho_E\left(\vec{r_1}\right)$ that controls the decay of~$\psi(\vec{r})$ in the regions where $E < W$, as dictated by Agmon's inequality~\cite{Agmon1982,Agmon1985}.  This distance is defined as
\begin{equation}
\rho_E\left(\vec{r_1},\vec{r_2}\right) = \min_{\gamma} \left(\int_\gamma \sqrt{\left( W\left(\vec{r}\right) - E \right)_+}~ds\right)
\end{equation}
where the minimum is computed on all paths~$\gamma$ going from $\vec{r_1}$ to~$\vec{r_2}$. The control on the amplitude $\psi(\vec{r})$ of an eigenfunction centered in $\vec{r_0}$ of energy $E$ is expressed through the inequality
\begin{equation}
\label{eq:Agmon_ineq}
| \psi\left(\vec{r}\right)| \lesssim~e^{\displaystyle -\rho_E(\vec{r_0},\vec{r})}~.
\end{equation}
This exponential decay can be observed in Fig.~\ref{fig:u1D} which displays the localization of quantum states in a random potential with periodic boundary conditions. The potential is piecewise constant on intervals of length~1, with values following a uniform law between 0 and $V_{\rm max}=8$, where units of $\hbar^2/2m$ are considered [see Fig.~\ref{fig:u1D}(a)]. The domain length $L=64$ thus corresponds to the total number of such intervals. Figure~\ref{fig:u1D}(b) displays the effective confining potential $W=1/u$ computed from solving~Eq.~\eqref{eq:udef}. This effective potential (the dashed line) is superimposed in Fig.~\ref{fig:u1D}(c) with the logarithmic plot of the third excited state ($\psi_3$). In a log scale, an exponential decay of the state amplitude translates into a linear drop. The wells of~$W$ for $\psi_3$ are the locations where $W < E_3$, and they are outlined in grey. One can observe that the decay of~$\psi_3$ occurs exactly at the barriers of~$W$ and stops across the wells. Fig.~\ref{fig:u1D}(d) presents a comparison between the amplitudes of $\psi_0$, $\psi_1$, and $\psi_3$ (the solid lines) and the estimates obtained using the Agmon distance build from~$W$ (dashed lines). We find that the amplitudes and their estimates are almost identical, and this result is robust for higher energies and for many realizations of the random potential.

\begin{figure}
\leftline{\raise1mm\hbox{\includegraphics[height=33mm]{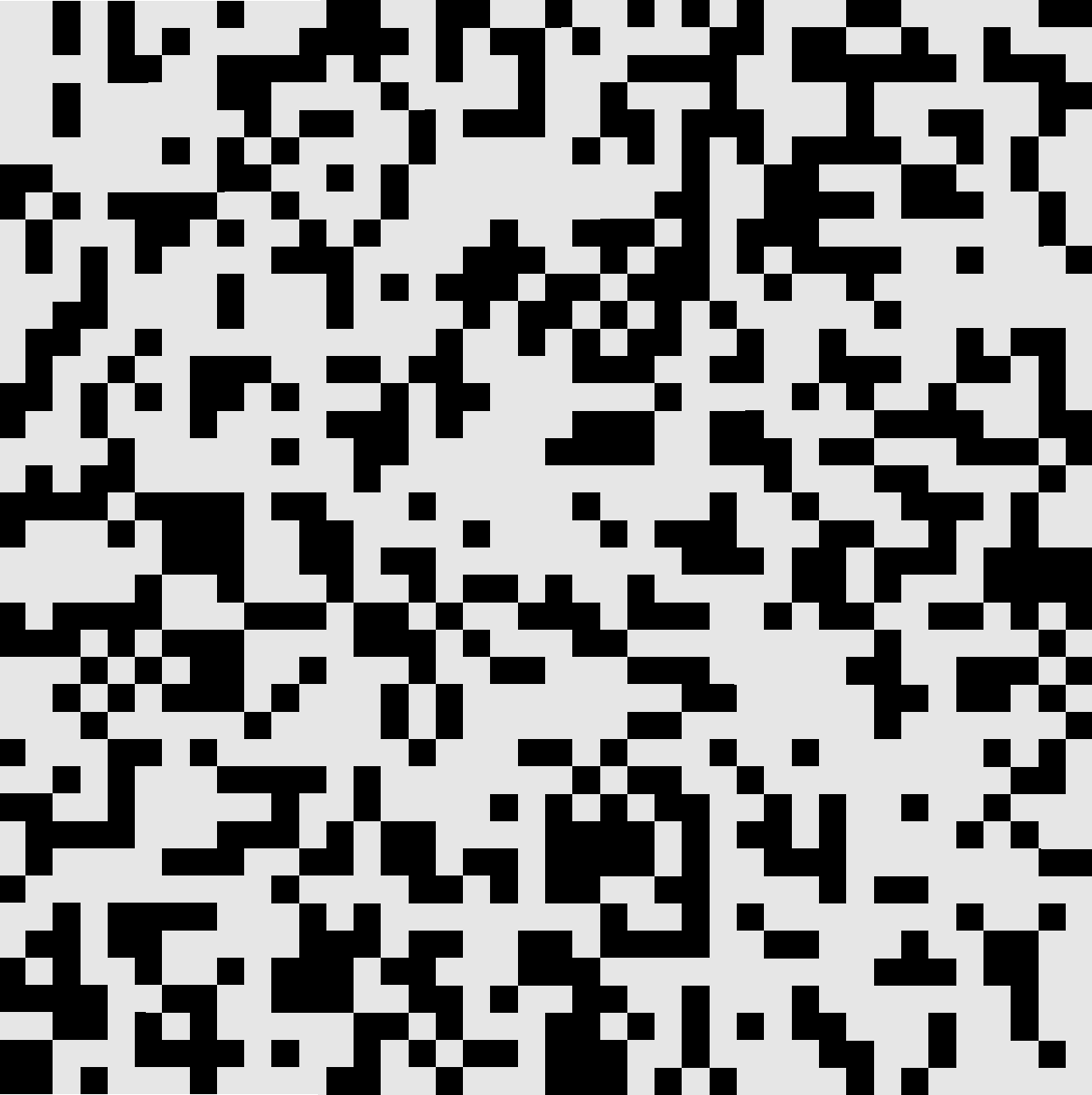}}%
\hspace{11mm}
\includegraphics[height=34mm]{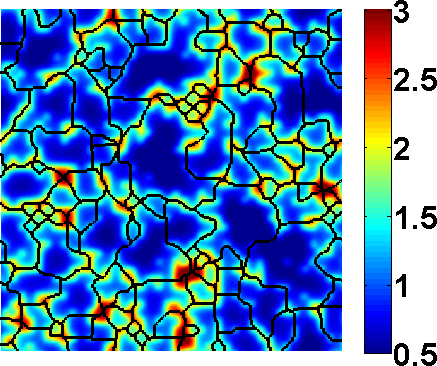}}
\leftline{\hspace{19mm}(a)\hspace{38mm}(b)}
\leftline{\includegraphics[height=34mm]{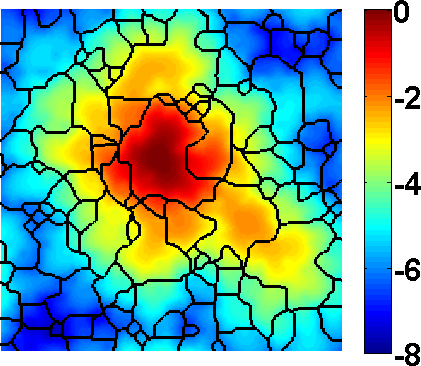}
\hspace{4mm}\hbox{\includegraphics[height=34mm]{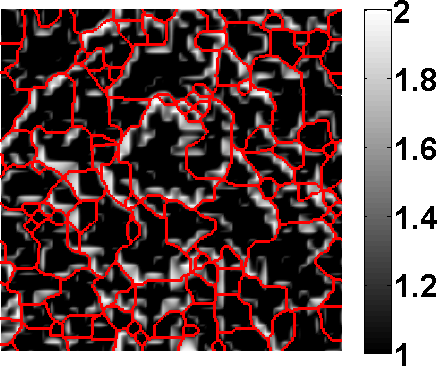}}}
\leftline{\hspace{19mm}(c)\hspace{38mm}(d)}
\caption{(a)~Boolean potential $V$.  The black region corresponds to $V=V_{max}=4$. The white region, corresponding to $V=0$, occupies about 60\% of the domain and percolates from the center to the outer boundary of the domain. (b)~Color representation of $W\equiv 1/u$, the effective potential as defined in Eq.~\eqref{eq:vp3}, superimposed by the crest lines of this potential computed using a watershed algorithm. (c)~Color representation of $\log_{10}[\psi_0(\vec{r})]$, $\psi_0$ being the fundamental eigenfunction of energy $E_0=0.503$, superimposed with the crest lines of $W$. (d)~The logarithmic gradient of $\psi_0$, thresholded so that values $>1$ are in black, and superimposed with the crest lines of $W$. The clear match indicates that the exponential decay of $\psi$ occurs exactly at the boundaries of the localization subregions determined by the effective potential~$W$.\label{fig:u2D}}
\end{figure}

The confining properties of $1/u$ are even more interesting in two dimensions. In the case of a 2D random potential where no clear localization region can be outlined, it has already been observed that the localization landscape~$u$ exhibits marked valleys that determine the localization subregions~\cite{Filoche2013}. Here we show that $1/u$ exponentially controls the confinement of the quantum states in the entire domain through the heights and widths of its barriers. This is of particular interest in cases where the semiclassical approach fails, as, for instance, for a Boolean-type potential (a random potential that can take only two values). Figure~\ref{fig:u2D}(a) displays a realization of a 2D~Boolean potential, i.e., a potential that can take only two values, $V_{\rm min}=0$ or $V_{\rm max}=4$. In the present computation, the domain is divided into $40 \times 40$ small unit squares, and the potential is piecewise constant on each of these unit squares, taking the value $V_{\rm min}$ with probability $p_0=0.6$ and the value $V_{\rm max}$ with probability $p=0.4$ so that the $V=0$ region has a large chance to percolate throughout the domain, as happens for the specific realization. Consequently, a classical particle would never be confined by such a potential, independent of its energy.

Both the effective potential $W$ and the fundamental quantum state $\psi_0$ are obtained using 320,000 triangular Lagrange finite elements of degree 3, with a code based on the FEniCS finite element software environment~\cite{LoggMardalEtAl2012a}. Figure~\ref{fig:u2D}(b) displays a color representation of~$W$, and its crest lines are computed using a watershed algorithm. Figure~\ref{fig:u2D}(c) presents the amplitude of $\psi_0$ overlapped with the same lines. One can notice how the decay of the amplitude closely follows the subregions delimited by these lines.

In order to examine in detail the long-range decay of this quantum state, the quantity $|\nabla\log(|\psi|)|$ which can be interpreted as the inverse of the local length of decay of the wave function, is plotted in Figure~\ref{fig:u2D}(d). Superimposing the level sets of this quantity with the crest lines of $1/u$ (which are also the valleys of the landscape~$u$) reveals a very strong correlation: the local length of decay increases substantially in the vicinity of the barriers of~$W$. The long-range exponential decay in Anderson localization does not occur uniformly but rather appears as the consequence of successive and cumulative decays across the dense network of barriers generated by~$W$. This behavior can be observed consistently in many trials and for all localized eigenfunctions.

In summary, the decay of the original eigenfunction $\psi=u~\vp$ away from its maximum thus originates from two concurring contributions:
\begin{itemize}
\item[(a)]
First, $u$ is small near the valleys surrounding the maximum of $\psi$, reducing accordingly the amplitude of~$\psi$ wherever $E~u\left(\vec{r}\right) < 1$~(see Ref.~\cite{Filoche2012}).

\item[(b)]
Second, $\vp$ decays through the barriers of $W$, which can be interpreted as ``quantum tunneling'' of the auxiliary function when $W$ is larger than~$E$.
\end{itemize}
So, not only does the amplitude of the quantum state~$\psi$ decrease in the vicinity of the valleys of the landscape~$u$, but it also decays when crossing the faraway valleys, as long as the value of $W$ is larger than the energy~$E$. This decay is directly governed by the exponential of the Agmon distance build from~$W$.

One can therefore conclude that the complicated interferential pattern of wave localization in~$V$ is translated into the picture of a classical confinement, observed through the glass of the wells and the barriers of the new effective potential~$W$. It is thus possible to graphically identify the localization subregions at energy~$E$ just by flooding the effective potential~$W$ up to the height~$E$, and by then observing the extent of the flooded basins. Each basin can be considered a local oscillator, the entire system then appearing as a set of oscillators, each independent at low energy and weakly coupled by quantum tunneling through the boundaries of the basins. {\color{black} One has to underline that for energies higher than the maximum of~$W$, there is no more classical confinement by $W$, and the state localization in 1D or 2D comes from the randomness present in the effective potential.}

It is interesting to note that $W$ can also be interpreted as a regularized version of the original potential~$V$. This can be seen by rewriting Eq.~\ref{eq:udef} as 
\begin{equation}
V - W = \frac{\hbar^2}{2m} \frac{\Delta u}{u}~.
\end{equation}
However, unlike in a classical smoothing procedure, the smoothing scale here is not constant but varies spatially depending on the value of $u$, and hence of $V$. The smoothed effective potential~$W$ resulting from this nonlinear operation favors the emergence of well-formed wells surrounded by barriers, even in situations where none are visible in the original potential. In that sense, the behavior of $W$ is much closer to that of a classical confining potential than~$V$.

The emergence of localized states triggered by the quenched disorder may also strongly perturb the one-electron density of states~\cite{Xu2010,Falco2010,Chen2012}. In some cases, it may shift the energy of the fundamental state, leading to an enhancement of the already existing gap in crystalline semiconductors~\cite{Thouless1974} or amorphous semiconductors~\cite{Mott1970} or the creation of a pseudogap in superconducting materials. Very generally, Weyl's law states that the integrated density of states (IDOS) $N(E)$ at a given energy~$E$ can be approximated by the volume (properly normalized) in the phase space~$\left(\vec{x},\vec{k}\right)$ that can be explored by a classical particle of mechanical energy smaller than~$E$:
\begin{align}
\label{eq:Weyl}
N(E) &\approx \left(2\pi\right)^{-n} \iint_{H(\vec{x},\vec{k}) \le E}~d^nx~d^nk~\nonumber\\&= \left(2\pi\right)^{-n} \iint_{\frac{\hbar^2k^2}{2m} + V(\vec{x}) \le E} d^nx~d^nk~.
\end{align}
where $n$ is the spatial dimension. For example in the local band structure theory of semiconductors, the local density of states is obtained by assigning to $V$ the value $E_c(\vec{x})$, the bottom edge of the conduction band. However, Weyl's formula is only valid in the asymptotic limit $E \rightarrow +\infty$ and can be very inaccurate at low energies.

It follows immediately from Eq.~\eqref{eq:Weyl} that Weyl's formula in one dimension writes $N(E)\approx N_V(E)$ where
\begin{align}
N_V(E) &= \frac{1}{2\pi}~\iint_{\frac{\hbar^2 k^2}{2m} + V(x) \le E}~dx~dk \nonumber\\& = \frac{1}{2\pi}~\int \left[\int_{\frac{\hbar^2 k^2}{2m} \le E - V(x)}~dk\right] dx \nonumber\\ &= \frac{\sqrt{2m}}{\pi\hbar}~\int_{V(x) < E} \left[E-V(x)\right]^{1/2} dx \label{eq:Weyl_V}~.
\end{align}

\begin{figure}
\begin{center}
\leftline{\raise3mm\hbox{
\includegraphics[width=0.23\textwidth]{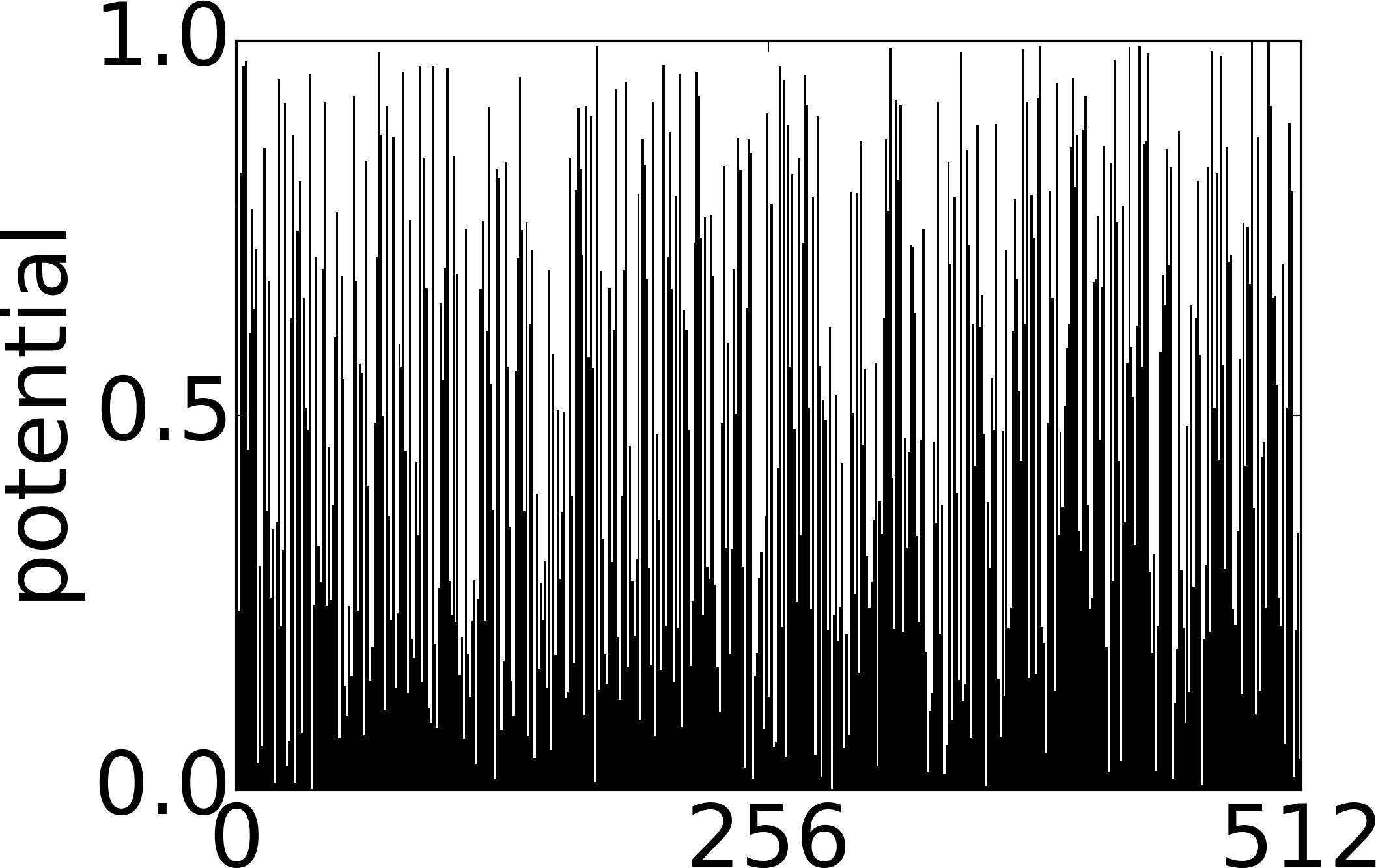}}
\includegraphics[width=0.23\textwidth]{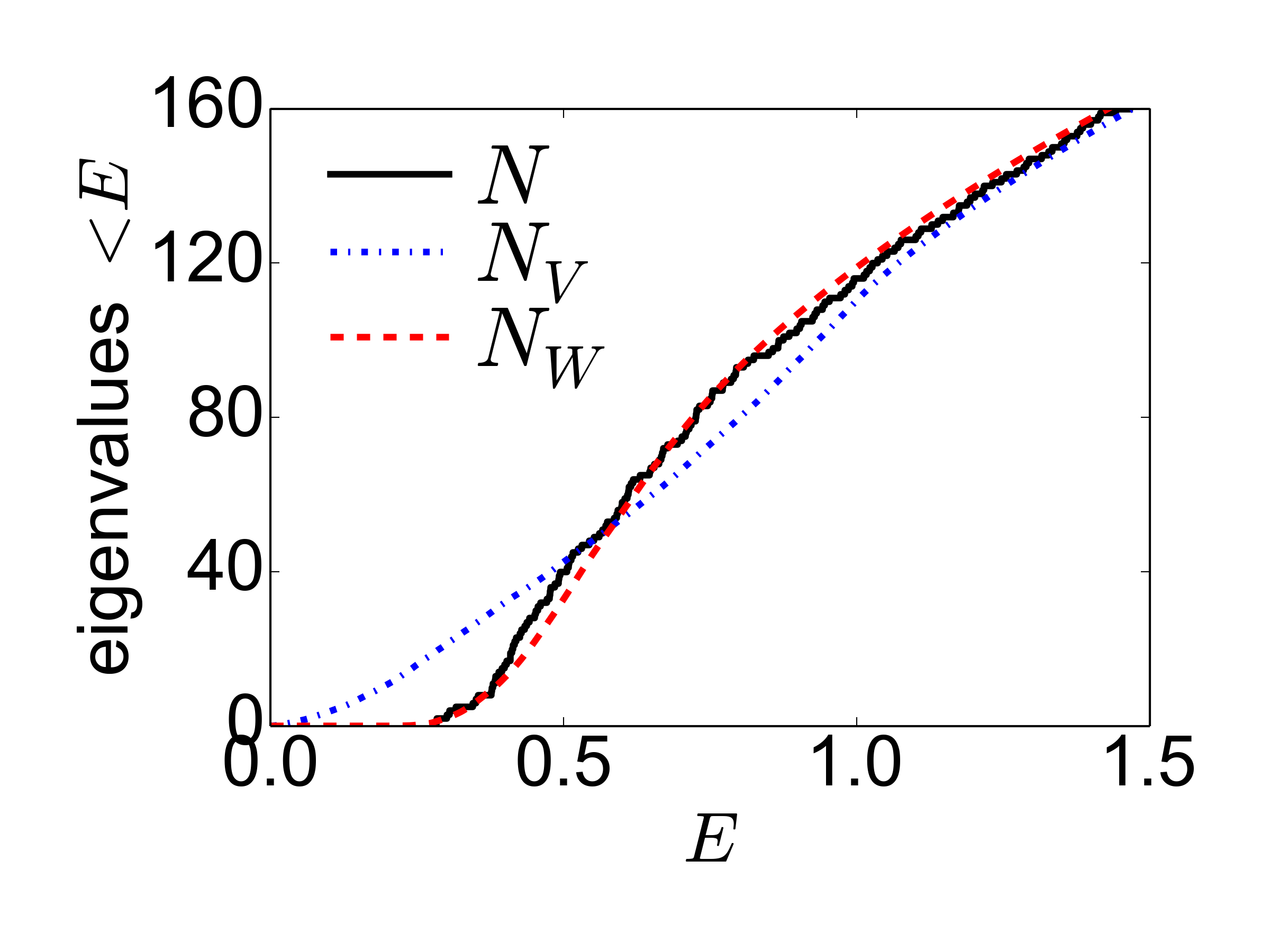}}
\leftline{\hspace{22mm}(a)\hspace{39mm}(b)}
\leftline{\raise3mm\hbox{
\includegraphics[width=0.23\textwidth]{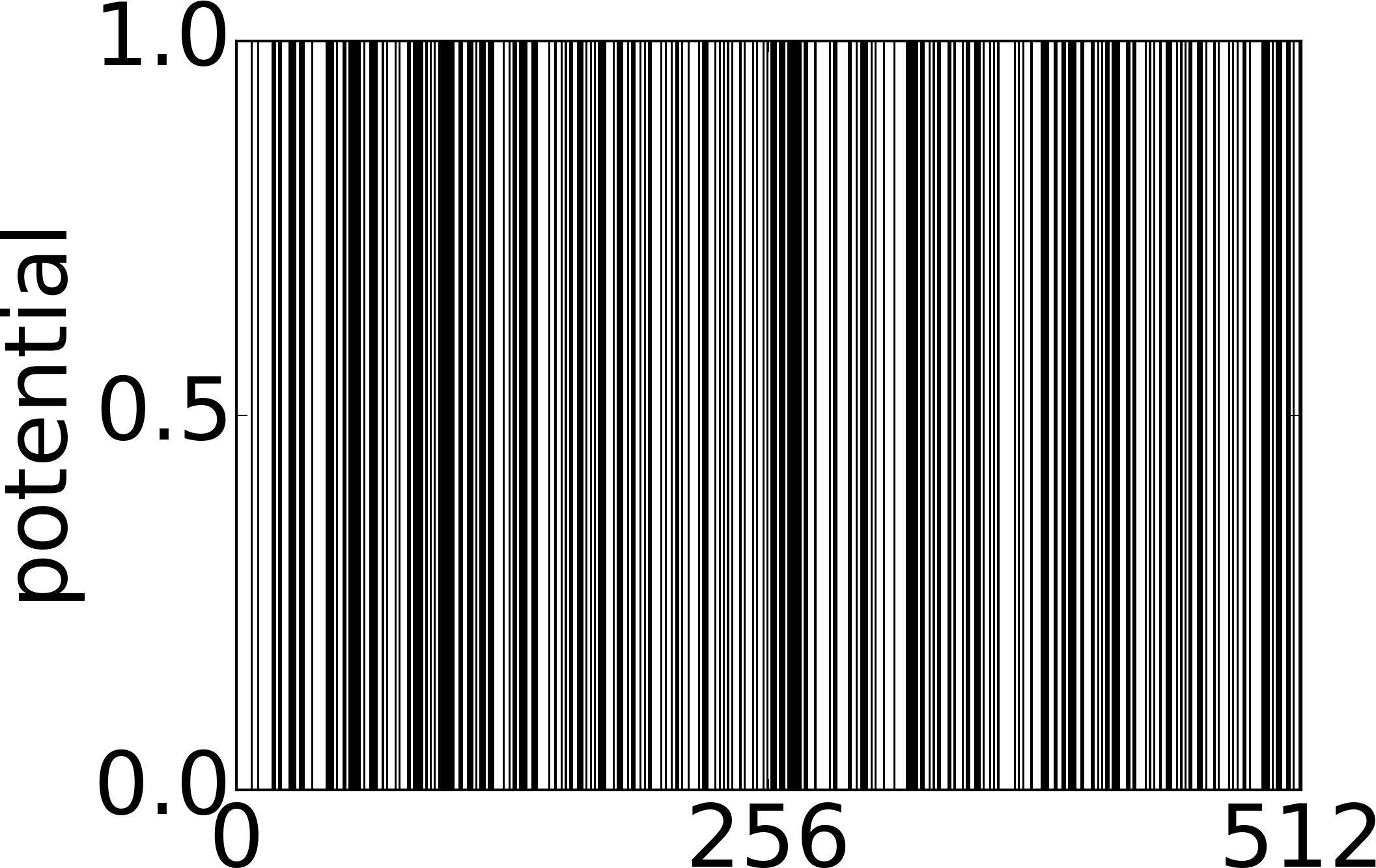}}
\includegraphics[width=0.23\textwidth]{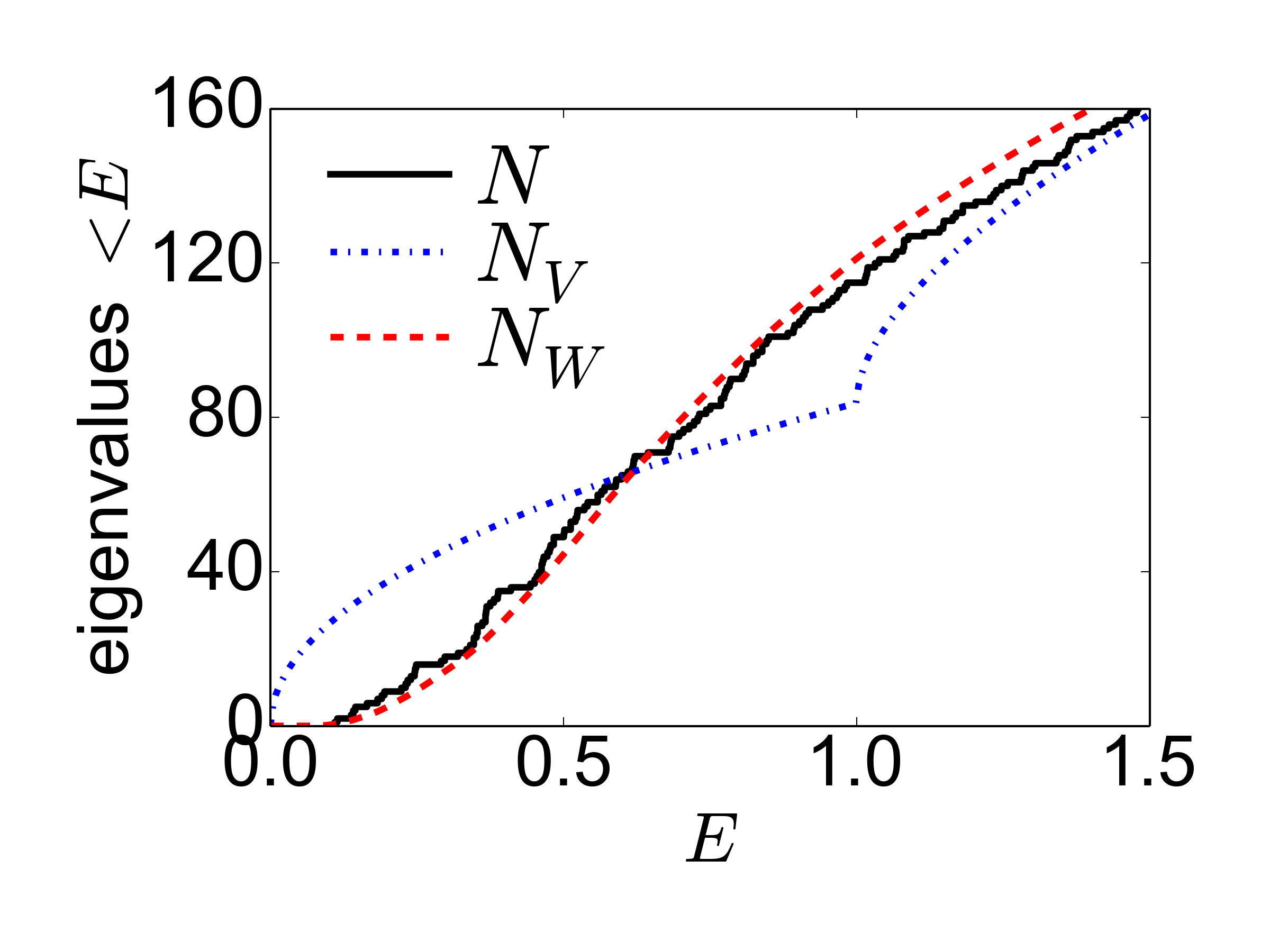}}
\leftline{\hspace{22mm}(c)\hspace{39mm}(d)}
\leftline{\raise3mm\hbox{
\includegraphics[width=0.23\textwidth]{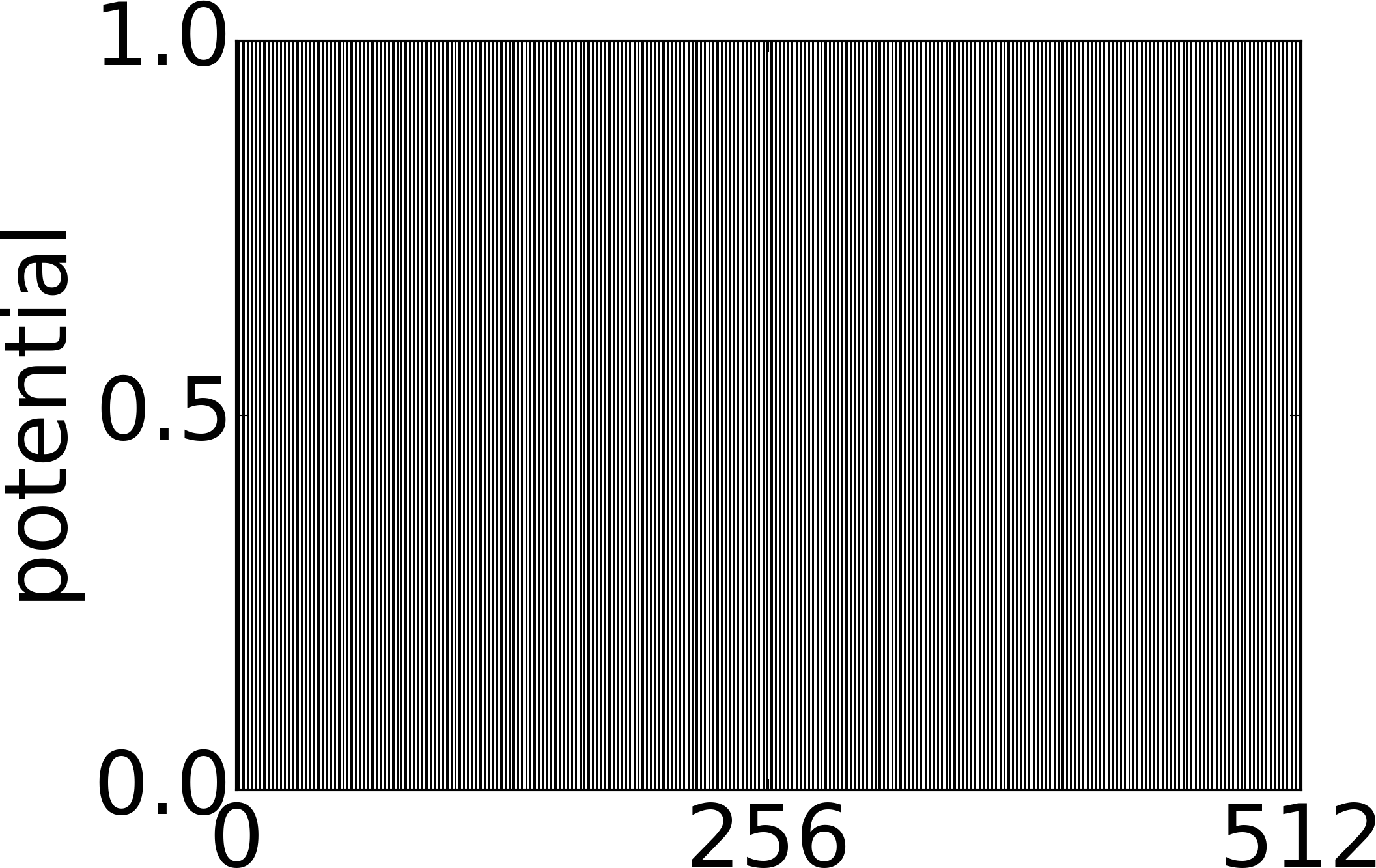}}
\includegraphics[width=0.23\textwidth]{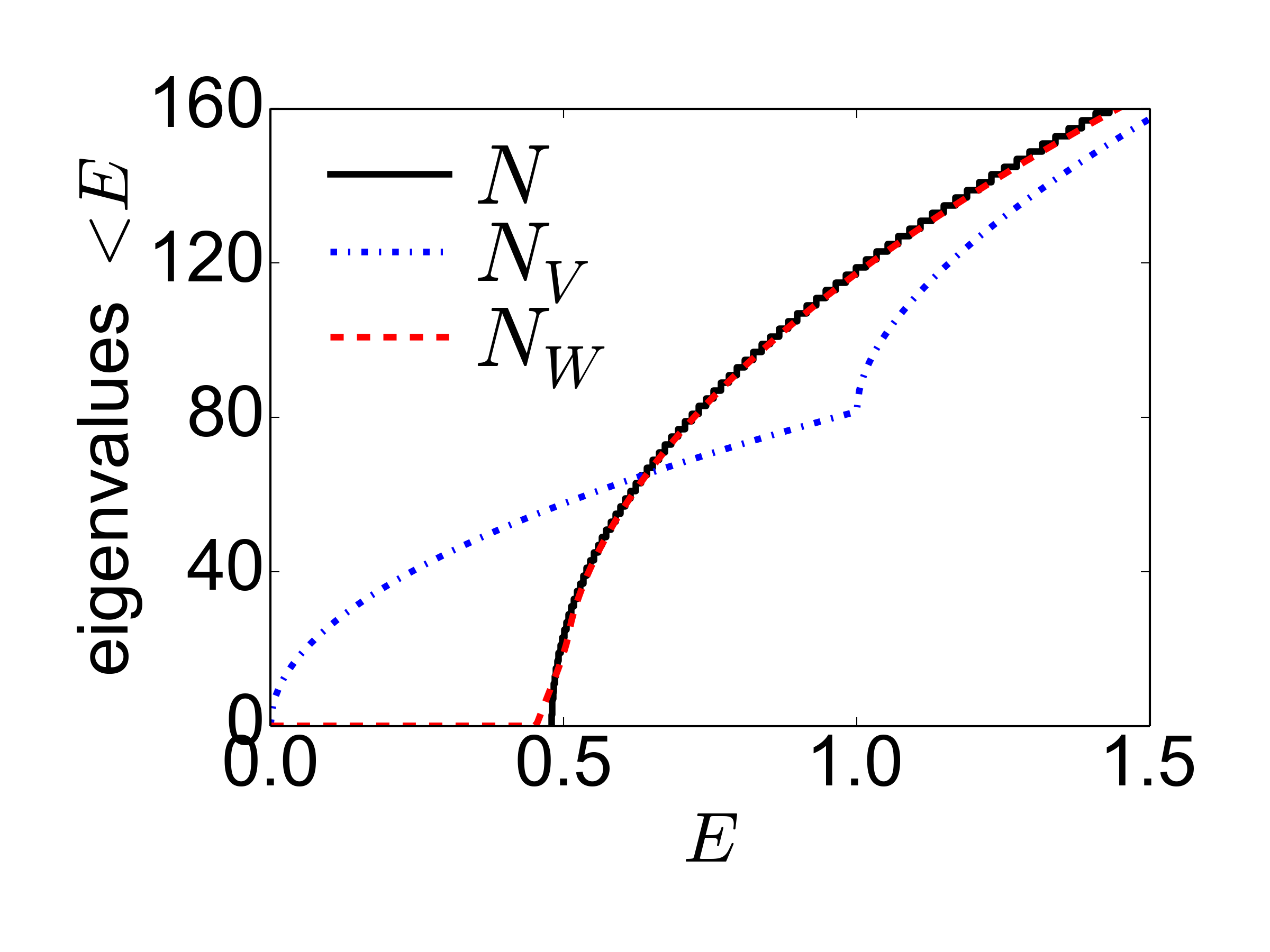}}
\leftline{\hspace{22mm}(e)\hspace{39mm}(f)}
\end{center}
\caption{(Left panels) Three different types of potentials: (a)~random with uniform law on [0 1], (c)~random Boolean (0 or 1), and (e)~periodic with 256 periods, $V=0$ on the first half of the period, $V=1$ on the second half. (Right panels) For each potential, the counting function $N$ (the solid black line) is represented, together with $N_V$ (the blue dash-dotted line), Weyl's approximation using the original potential $V$ [as defined in Eq.~\eqref{eq:Weyl_V}], and $N_W$ (the red dotted line), the same with $W$. Notice in all cases the remarkable agreement between $N$ and $N_W$.
\label{fig:Weyl}}
\end{figure}

Because of the analogy with classical mechanics that is implicitly behind Weyl's law, one can easily understand why this approximation is poor in quantum systems where wave interference plays a major role. This can be seen in Fig.~\ref{fig:Weyl} where the density of states for three typical cases of one-dimensional potential are examined: (a)~random with uniform law, (c)~random boolean, and (e)~periodic. For each type of potential represented on the left, both the true counting function $N(E)$ (the stepwise solid line) and the approximation $N_V(E)$ (the dotted red curve, obtained from Weyl's approximation) are represented. One can see that in all three cases, these two curves differ significantly as expected.

We now compute a new approximation of the counting function, this time based on the potential~$W$ deduced from~$u$. The reason for doing so lies in the fact that $W$ has been shown to behave as an effective potential energy [see~Eq.~\eqref{eq:Agmon}] with identifiable wells and barriers, in the spirit of a classical mechanical system. In all three of the frames (b), (d), and (f), the smooth curve (the dotted red line) is the approximation $N_W(E)$ obtained by inserting the effective potential~$W$ into Eq.~\eqref{eq:Weyl_V}. One can immediately notice the remarkable agreement between the original counting function and $N_W(E)$, whereas the standard approximation~$N_V$ using the original potential totally fails to predict the density of states. In particular, $N_W(E)$ detects precisely the shift of the lower edge of the conduction band induced either by disorder~[see Fig.~\ref{fig:Weyl}(b)] or by the periodicity of the potential [see~Fig.~\ref{fig:Weyl}(f)]. This lower bound to all energy values corresponds to an added gap in the IDOS that is reminiscent of the missing density of states at lower energy in Anderson localization~\cite{Mott1969}. At higher energy, all three curves catch up, as they asymptotically follow Weyl's law without potential, which is proportional to~$E^{1/2}$.

In summary, a new conceptual tool has been introduced for understanding the localization properties of quantum states. This object is the reciprocal of the localization landscape introduced in Ref.~\cite{Filoche2012} and can be interpreted as an effective potential. We have shown that this effective potential not only determines the boundaries of the localization regions but also controls the long-range decay of the quantum states through an Agmon metric. In particular, the exponential decay observed in Anderson localization has been shown not to occur uniformly, but rather to be concentrated across the barriers of this effective potential. Therefore, the transport between adjacent basins of the effective potential involves an ``effective quantum tunneling.'' In short, one can say that the effective potential captures the interference pattern created by waves in the original potential and converts it into a semiclassical picture of confining potential. Finally, this property enables us to build an approximation of the integrated density of states based on Weyl's law. Applied to various types of 1D potentials, either random or deterministic, this approximation showed a remarkable agreement with the actual IDOS, far more accurate than what can be obtained using Weyl's law, with or without the original potential.

Because it captures the deep localization properties of a complex or random potential, the effective confining potential is a very promising tool for understanding the features of quantum waves in disordered media. Indeed, our procedure reveals the system as a partition of weakly coupled oscillators, and it thus efficiently realizes an approximate diagonalization of the Hamiltonian for the portion of the spectrum spanned by localized eigenfunctions. With this information at hand, we can calculate the carrier distribution in random alloys such as, for instance, the active layers of GaN-based Quantum-well light-emitting diodes~\cite{Criproni}. Finally, our approach brings a new perspective to such open questions as the nature of the transition from localized to delocalized states in three dimensions (the mobility edge) and the onset of localization in many body systems.

\begin{acknowledgments}
D.~N.~A.\ is partially supported by the U.S.~NSF Grant No.~DMS-1418805. G.~D.\ is member of the Institut Universitaire de France, and is partially supported by an ANR Grant, programme blanc GEOMETRYA, Grant No.~ANR-12-BS01-0014. D.~J.\ is partially supported NSF Grants No.~DMS-1069225 and No.~DMS-1500771. S.~M.\ is partially supported by the Alfred P.\ Sloan Fellowship, NSF CAREER Grant No.~DMS-1056004, the NSF MRSEC Seed Grant, and the NSF INSPIRE Grant No.~DMS-1344235. M.~F.\ is partially supported by a PEPS-PTI grant from CNRS.
\end{acknowledgments}

\bibliography{localization}

\end{document}